\documentclass{article}

\pdfoutput=1
\usepackage{arxiv}
\usepackage[utf8]{inputenc} % allow utf-8 input
\usepackage[T1]{fontenc}    % use 8-bit T1 fonts
\usepackage{amsfonts}       % blackboard math symbols
\usepackage[intlimits]{amsmath}
\usepackage{nicefrac}       % compact symbols for 1/2, etc.
\usepackage{microtype}      % microtypography
\usepackage{graphicx}
\usepackage{doi}

\usepackage{chngcntr}

\usepackage{lineno}
\usepackage{float}
\usepackage{cleveref}

\usepackage{tikz}
\newcommand*\circled[1]{\tikz[baseline=(char.base)]{
            \node[shape=circle,draw,inner sep=0.5pt] (char) {#1};}}

\title{Fabrication and transfer print based integration of free-standing GaN membrane micro-lenses onto semiconductor chips}

\author{Nils Kolja Wessling \\
    \textbf{Benoit Guilhabert}\\
    \textbf{Miles Toon}\\
    \textbf{Martin D.\,Dawson}\\
    \textbf{Michael J. Strain}\\
	Institute of Photonics\\
	Department of Physics\\
	University of Strathclyde \\
    Glasgow, UK\\
    \textit{michael.strain@strath.ac.uk}
	\And
	Saptarsi Ghosh\\
    \textbf{Menno Kappers}\\
    \textbf{Rachel\,A.\,Oliver}\\
	Cambridge Center for Gallium Nitride\\
    University of Cambridge\\
    Cambridge, UK
}

\renewcommand{\headeright}{}
\renewcommand{\undertitle}{}
\renewcommand{\shorttitle}{Fabrication and transfer print based integration of free-standing GaN membrane micro-lenses onto semiconductor chips}

\hypersetup{
pdftitle={Fabrication and transfer print based integration of free-standing GaN membrane micro-lenses onto semiconductor chips},
pdfsubject={physics.app-ph},
pdfauthor={Nils Kolja Wessling, Saptarsi Ghosh, Benoit Guilhabert,Menno Kappers,Rachel A. Oliver, Martin D. Dawson, Michael J. Strain},
pdfkeywords={Additive micro-optics, GaN, micro-lens, diamond, integrated photonic circuits, transfer printing}
}

\begin{document}
\maketitle

\begin{abstract}
	We demonstrate the back-end integration of broadband, high-NA GaN micro-lenses by micro-assembly onto non-native semiconductor substrates. We developed a highly parallel micro-fabrication process flow to suspend micron scale plano-convex lens platelets from 6" Si growth wafers and show their subsequent transfer-printing integration. A growth process targeted at producing unbowed epitaxial wafers was combined with optimisation of the etching volume in order to produce flat devices for printing. Lens structures were fabricated with $6-11$\,µm diameter, 2\,µm height and root-mean-squared surface roughness below 2\,nm. The lenses were printed in a vertically coupled geometry on a single crystalline diamond substrate and with $\mu$m-precise placement on a horizontally coupled  photonic integrated circuit waveguide facet. Optical performance analysis shows that these lenses could be used to couple to diamond nitrogen vacancy centres at micron scale depths and demonstrates their potential for visible to infrared light-coupling applications.
\end{abstract}

% keywords can be removed

\section{Introduction}
Micron-sized optical elements are important components for the coupling of light between free-space and matter based systems. Dielectric micro-lens arrays (MLA) are commercially available and commonly used to focus light onto the photon-sensitive regions of CCD\,\cite{Stevens2010} and CMOS\,\cite{DouglasA.Baillie2004} photo-detector arrays, in order to increase quantum efficiency. Micro-optics are also used to improve the efficiency of light extraction from light emitting diodes (LEDs)\,\cite{Braeuer2003} and $\mu$-LEDs\,\cite{Choi_2_2005}, for coupling to integrated photonic circuits (PICs)\,\cite{Marchetti2019} as well as for improving pump and collection efficiency between microscope systems and semiconductor based solid state quantum emitters\,\cite{Bogucki2020}. This latter application is particularly challenging due to the large refractive index of the host material that leads to total internal reflection at the interface between the material and free-space.\\

There are two main routes to fabrication of micro-optics on semiconductor substrates, either additive materials patterning on the substrate, or direct fabrication of optical surfaces into the material itself. There are a number of techniques available for the fabrication of additive components, including imprint lithography\,\cite{Voelkel2012,Moharana2020}, two-photon polymerization\,\cite{Gissibl2016,Dietrich2018,Gehring2019}, thermal reflow\,\cite{Oder2003,Choi2005,Sardi2020}, and inkjet printing\,\cite{Zolfaghari2019,Fischer2020}. Structures formed by these methods are commonly used as final optical element themselves, but may also be used as a mask for a dry-etch shape transfer into the semiconductor. Two photon polymerization (TPP) is a particularly flexible method and can be used to fabricate complex micro-optical systems such as multi-lens beam expanders on fibre tips\,\cite{Gissibl2016} as well as vertical and end-fire couplers to PICs\,\cite{Dietrich2018,Gehring2019}. The relatively low refractive index of TPP resins (1.5-1.6\,\cite{NanoScribeGmbH&Co.KG2022} compared with common semiconductor materials >2.0\,\cite{Green2008,Xu2014,Phillip1964,Barker1973,Pastrnak1966}) limits the achievable numerical aperture\,(NA) of the polymer lenses.  Furthermore, polymeric materials are susceptible to catastrophic optical damage at watt-level optical power\,\cite{Dietrich2018}.\\
Alternatively, solid immersion lenses (SIL) can be formed directly into semiconductor materials by focused ion-beam milling\,\cite{Marseglia2011,Jamali2014}, laser-micromachining\,\cite{Presby1990, Hao2012}, photoresist reflow in combination with reactive ion etching\,\cite{Gu2004,Lee2008}, dual masking\,\cite{Zhang2017} and diamond turning\,\cite{To2016,Mukaida2017}. These methods make use of the high refractive index of the material itself to avoid refractive index contrast at interface layers, produce high NA performance and are robust to optical damage. Individual lenses can be fabricated with high positional accuracy but require either serial, time consuming processing\,\cite{Marseglia2011,Jamali2014}, or are in the case of diamond substrates limited in the form factors that can be achieved due to the mask-to-semiconductor etch selectivity\,\cite{Gu2004,Lee2008}.

\begin{figure}[h]
    \centering
    \includegraphics[width=\linewidth]{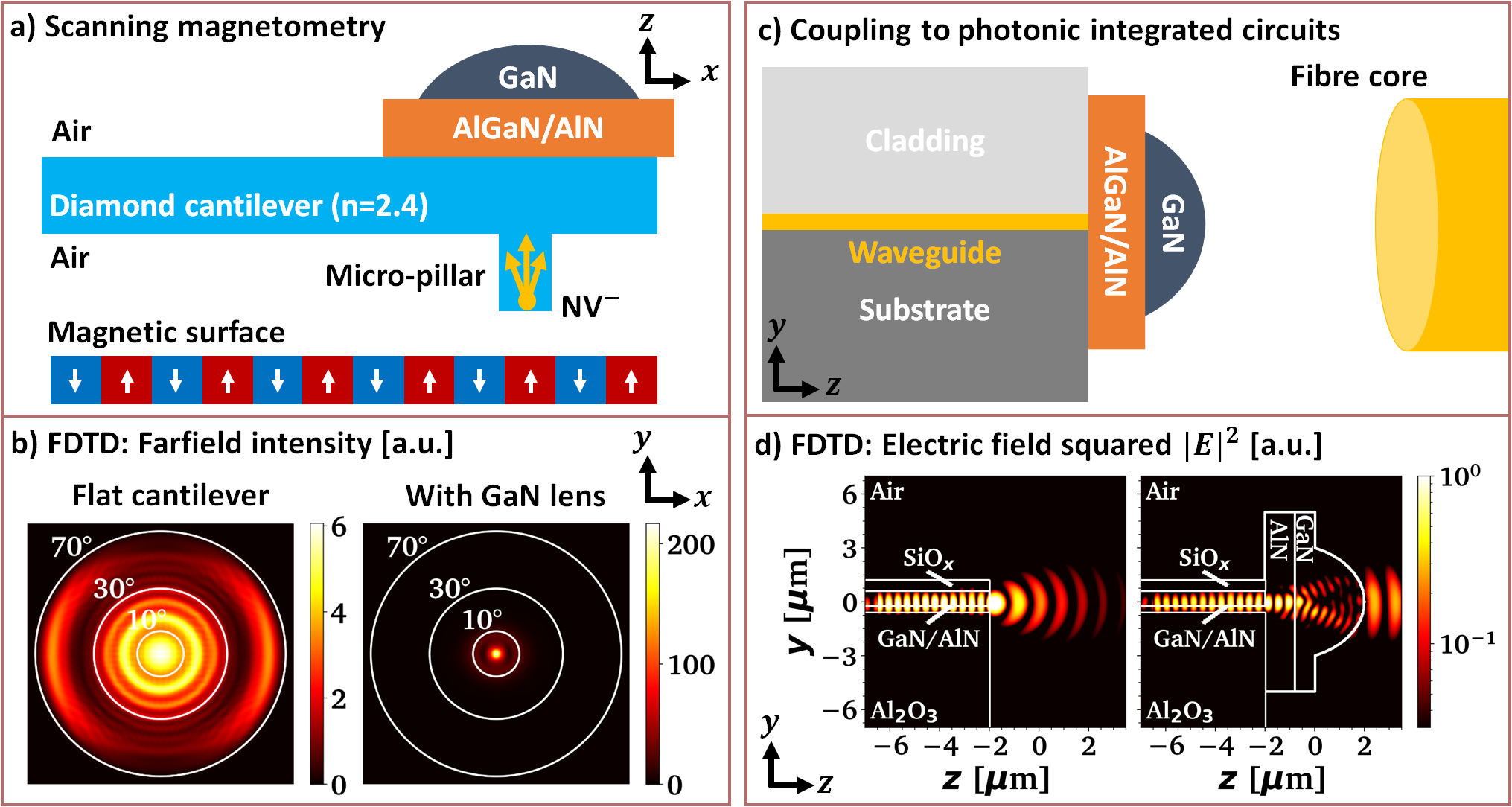}
    
\caption{a)\,Application example for diamond NV scanning magnetometry, exploiting a GaN lens as a light coupler, b)\,Simulated far field of a dipole emitter placed in a 200 nm wide square micro-pillar (10 nm away from the pillar tip, assuming [111]-orientation) without and with a printed GaN micro-lens (ROC\,$=16\,\mu$m, 15\,$\mu$m distance to the top of the pillar) at $\lambda=650$\,nm wavelength c)\,Schematic of a GaN micro-lens facilitating end-fire coupling between a photonic integrated circuit (left) and a fibre core (right), d)\,Simulated free-space divergence from the fundamental TE mode of an GaN/AlN waveguide (1.2\,$\mu$m height, 2\,$\mu$m width) without and with a GaN micro-lens (ROC\,$=3.3\,\mu$m) printed on the waveguide facet.}
\label{fig:Motivation}
\end{figure}

In this work, we present an alternative fabrication scheme, whereby micro-lenses are directly fabricated in GaN material and subsequently transferred to a host chip as membrane devices. The use of a III-N material enables better etch selectivity than in e.g. diamond, while retaining a refractive index match to semiconductor material. The devices are fabricated using a combination of grayscale lithography  and the photoresist reflow to create spherical photoresist micro-lenses, followed by inductively coupled reactive ion etching (ICP-RIE) to transfer the pattern into a GaN/AlGaN/AlN layer stack grown on 6" Si wafers\,\cite{Ghosh2021}. The micro-lenses are suspended over the silicon substrate by wet chemical etching and transferred to non-native substrates using a transfer print process. We demonstrate transferred micro-lenses with micron scale radius of curvature and focal length on diamond substrates. We also show direct printing of a micro-lens onto the facet of a GaN waveguide chip. Two example applications are shown in \figurename\,\ref{fig:Motivation}. In\,a), light is coupled to and from a diamond nitrogen vacancy (NV$^-$) centre into free-space collection optics for scanning magnetometry\,\cite{Balasubramanian2008,Maletinsky2012}.  By including a GaN micro-lens on the diamond substrate, the effective collection angle from the device is significantly reduced, improving potential coupling efficiency with external, low NA optics. \figurename\,\ref{fig:Motivation}\,b) shows a scenario where a micro-lens is integrated onto the facet of a GaN/AlN planar waveguide chip, again improving modal coupling efficiency between the highly confined on-chip mode and low NA external optics such as a single mode fiber.

\section{Fabrication and device transfer}

\begin{figure}[h]
\centering\includegraphics[width=\linewidth]{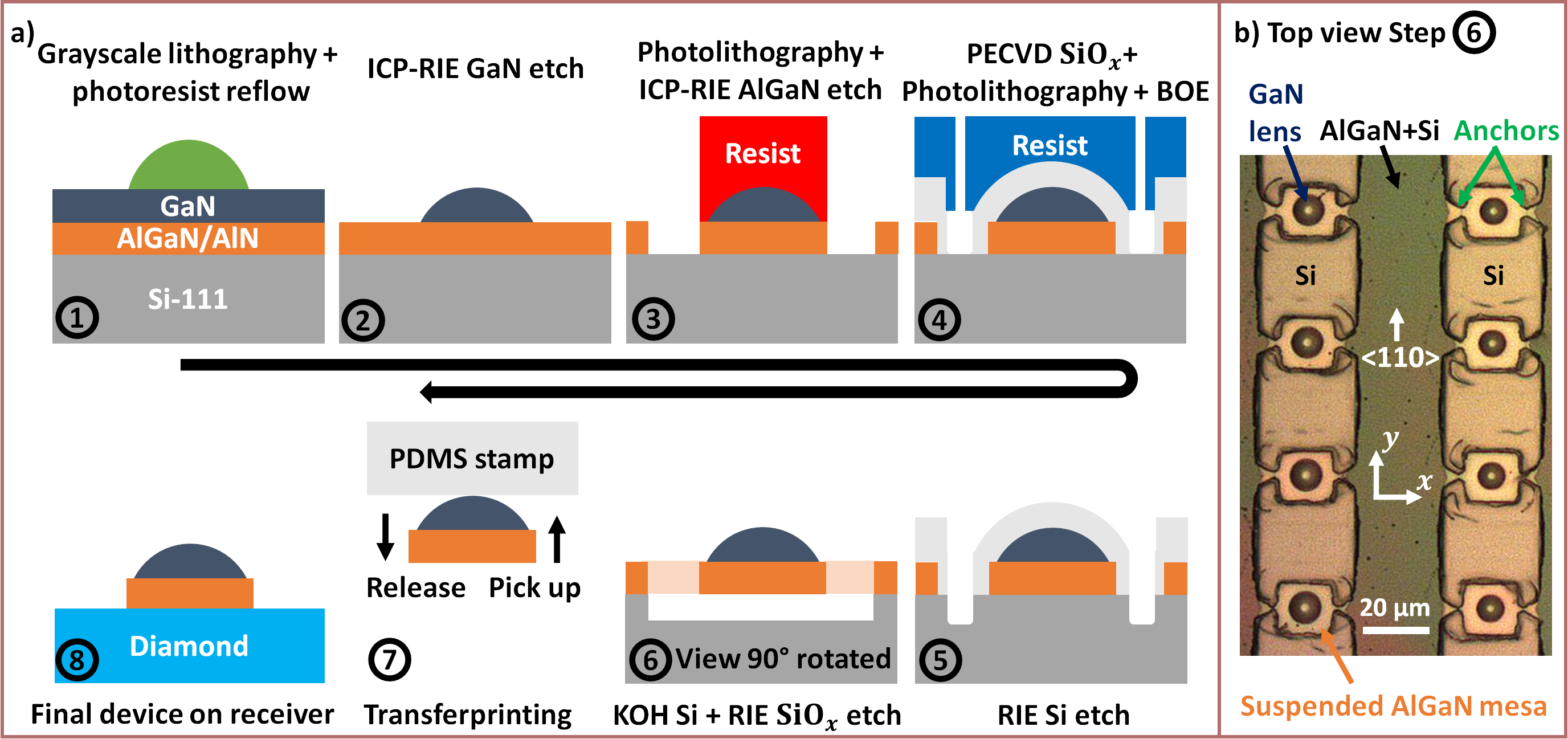}
\caption{a)\,Illustration of the process flow after growth, using CVD-grown, single crystalline diamond as receiver substrate, b)\,Microscope image of GaN-on-Si donor chip corresponding to process step 6.}
\label{fig:ProcessFlow}
\end{figure}

Fig.\,\ref{fig:ProcessFlow} illustrates the process flow for GaN micro-lens fabrication and transfer. A wafer die is spin coated with "Shipley SPR220-4.5" photoresist and micro-lenses are defined using grayscale laser lithography (Heidelberg Instruments DWL66+)\,\circled{1}. To smooth the surface and form a spherical lens profile with the target height, a 130-150\,°C reflow bake is applied to the sample on a hotplate. An Ar/Cl$_2$-based ICP-RIE (200\,W coil power, 70\,W platen power, 30\,sccm Cl$_2$, 10\,sccm Ar, 20\,mTorr) is used to transfer the lens shape into the GaN \,\circled{2}. The ICP recipe yields etch rates of 200\,$\frac{\text{nm}}{\text{min}}$ for SPR220-4.5 and 150\,$\frac{\text{nm}}{\text{min}}$ for GaN/AlGaN, leading to a selectivity around $0.75$. The etch rate drops to around 120\,$\frac{\text{nm}}{\text{min}}$ for AlGaN with Al content $x>0.6$ and in the AlN layer. The reactive ion etching results in a lens profile transformation from a spherical resist lens to a parabolic shape of the etched lens, as previously observed in SiC\,\cite{Sardi2020}, indicating a partly chemical etch and possibly ion channeling effects.\\
The defined GaN lenses are overlaid with a mesa pattern including suspension anchors using "Shipley SPR220-7.0" photoresist. The previously detailed ICP-RIE recipe is employed to remove the remaining AlGaN and AlN layers, leading to a slight overetch into Si\,\circled{3}. Fig.\,\ref{fig:ProcessFlow}\,b) shows a microscope image of already suspended devices, illustrating the etched anchor and mesa pattern defined in this etching step. The 7\,$\mu$m thick photoresist is needed to successfully protect the GaN lens surface from the plasma.

After resist removal, a conformal layer of 1.8\,$\mu$m thick SiO$_x$ is deposited by plasma enhanced chemical vapor deposition (PECVD). This hard mask protects the GaN lens from the KOH wet etch in step \circled{6}. SPR220-7.0 is used as a lithography mask to open windows at the bottom of the etched trenches and a buffered oxide etch (7:1) removes the SiO$_x$ with a slight undercut, leaving a sidewall protection layer of SiO$_x$ on the AlGaN/AlN mesa below the lens\,\circled{4}.\\
An anisotropic RIE etch (100\,sccm SF$_6$, 8\,sccm O$_2$, 25\,mTorr, 50\,W RF-power\,\cite{Shaw1994}) is applied after resist removal to etch the Si substrate using the SiO$_x$ layer as a mask\,\circled{5}. The resultant trench allows access for a KOH wet etch solution (40\,\% in weight, 85\,°C) to the (110) crystal plane, selectively removing the Si below the lens-mesa along the fast <110> etch direction\,\circled{6}. The schematic in Fig.\,\ref{fig:ProcessFlow} is rotated for this process step, indicating the anchors as semitransparent areas and showing the anisotropic nature of the KOH etch in Si(111). Fig.\,\ref{fig:ProcessFlow}\,b) shows a top view. As previously confirmed by scanning transmission electron microscopy\,\cite{fairclough2021direct,Ghosh2021}, a nm-thin disordered AlSi$_y$N$_x$ interlayer is formed between Si substrate and AlN nucleation layer, which provides an effective bottom protection against the KOH solution. The smoothness of the mesa bottom surface is confirmed by atomic force microscopy (AFM) measurements after flipping a flat membrane device with a PDMS stamp, yielding 0.4\,nm root-mean-squared (r.m.s.) roughness.\\
The SiO$_x$ layer is removed by RIE dry etching (5\,sccm CHF$_3$, 15\,sccm Ar, 30\,mTorr, 120\,W RF-power) to restore the GaN lens surface. AFM surface roughness analysis on a GaN micro-lens after SiO$_x$ removal showed a surface roughness on the order of 1.4\,nm r.m.s.\\
The suspended micro-lenses are transfer-printed to receiver substrates (single crystalline\,(SC) diamond or GaN/AlN waveguide facets) using a custom transfer print system with sub-micron spatial precision\,\cite{McPhillimy2020}.  The lenses are removed from their growth substrate using a soft polymer stamp fabricated using a 6:1 ratio Sylgard PDMS with a contact area of 30x30$\,\mu$m$^2$\,\circled{7}/\circled{8}. The devices are finally aligned to the receiver chip position and released, making use of Van-der-Waals interaction between the lens bottom and receiver chip top surface.\\
Additional data on the process flow, surface roughness, lens-shape transformation and process yield is given in the supplementary material Fig.\,\ref{fig:ProcessDetails}-\ref{fig:YieldofSuspension}.

\section{Material growth for membrane flatness}
The success of the heterogeneous integration scheme detailed above relies on the flatness of the released membrane devices in addition to the back-side smoothness. If the membrane devices are too bowed, it becomes impossible to achieve a good contact with the receiver substrate, leaving an air cavity between the substrate and the device surface, which in the case of micro-lens devices can lead to interface reflections and distortion of the desired lens behaviour.  Appropriately flat membranes are achieved via careful optimisation of the growth process for the GaN heteroepitaxial layers. The structures used for micro-lens fabrication in this work are grown by metal organic vapor phase epitaxy\,(MOVPE) on Si(111) substrates. As-grown, the multilayers consist of a 0.25$\,\mu$m AlN nucleation layer, a 1.7\,$\mu$m graded Al$_x$Ga$_{1-x}$N buffer (with AlN mole fraction $x$ decreasing from 75\,\% to 25\,\%), and a 2\,$\mu$m GaN layer (i.e. a total thickness of ca. 4\,$\mu$m nitride material). By growing in succession layers with a decreasing Al content, each layer has a slightly larger relaxed lattice constant than that below it, so that the epitaxial growth induces compressive stress. This stress counteracts the post-growth cooldown related tensile stress originating from the thermal expansion mismatch between the silicon substrate and the nitride epitaxial layers. Post-growth, the wafer-bows are functions of the residual stresses and thickness of the individual epilayers. For membranes released from the growth substrates, their bows are also dependent to these in-built stresses in the remaining layers and related to the wafer bow\,\cite{Spiridon2021}.

\begin{figure}[h]
\centering\includegraphics[width=0.75\linewidth]{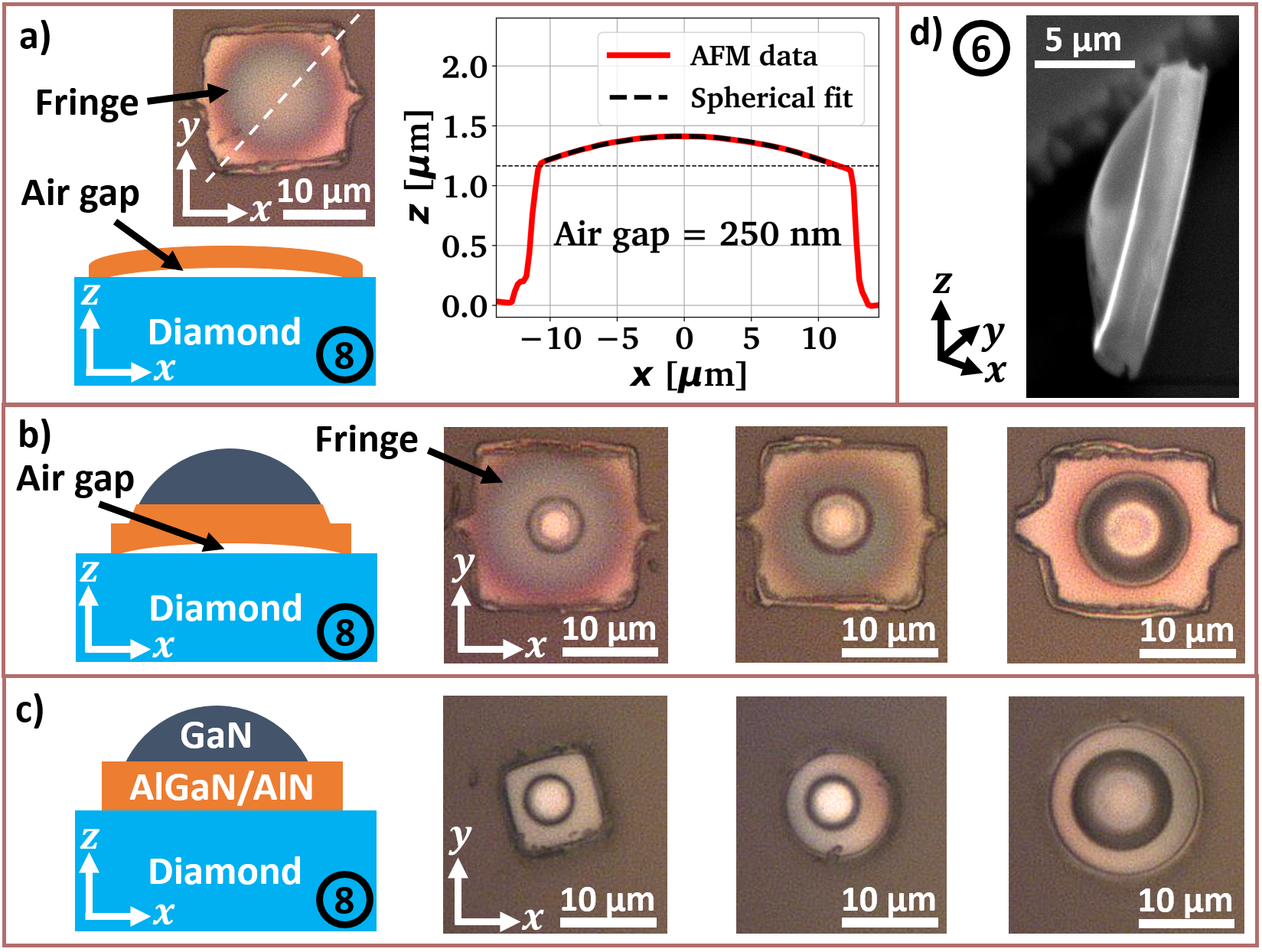}
\caption{a)\,Schematic, microscope image and AFM analysis of a transfer-printed 1.2 $\mu$m thin AlGaN/AlN membrane printed onto SC diamond, b)/c)\,Schematic and microscope images of $2-3$\,$\mu$m high micro-lenses in b) on bowed, $1.2\,\mu$m thick mesas with anchors and in c) on flat, 2\,$\mu$m thick mesas without anchors printed on SC diamond, d)\,40\,° tilted SEM image of an upright-standing 2\,$\mu$m high GaN lens with a flat, 2\,$\mu$m thick mesa on the donor substrate.}
\label{fig:MembraneBow}
\end{figure}

The wafer bow can be controlled by exposing the substrate to ammonia at 1000\,ºC prior to deposition of the AlN nucleation layer. By changing the duration of the ammonia pre-dose, the bow may be tuned from convex to concave. Initial trials  of the membrane fabrication process with the pre-dose duration varying (without any lens fabrication) used a series of wafers in which a 1\,$\mu$m GaN epilayer was grown on top of similar underlying buffer layers to those employed for the micro-lens fabrication. These trials indicated that minimising the wafer bow also yielded the least bowed membrane (see supplementary material, Fig.\,\ref{fig:ChipletBow}).  Based on this information, for the present studies with 2\,$\mu$m thick GaN layers, material from two nearly-flat wafers (having bow values of $+7\,\mu$m and $-2\,\mu$m) are used for the experiments. It is observed that even for material from flat wafers, the material removal in the micro-lens fabrication needs to be further optimised to realise flat devices.\\
For example, Fig.\,\ref{fig:MembraneBow}\,a) shows a test membrane with a planar surface (i.e. no lens topology) where material was removed by ICP-RIE until 1.2\,$\mu$m of the combined graded AlGaN buffer and AlN nucleation layer remained. Though the source wafer had near-zero bow, when the resulting AlGaN/AlN membrane was transferred to a diamond substrate, a clear interference fringe pattern appears. This indicates an air gap below the device. An AFM scan confirms convex bowing of the membrane, showing that the maximum width of the air gap is on the order of 250 nm.\\
Membranes with micro-lens topology are also bowed when etched to this same remaining thickness of 1.2\,$\mu$m, as shown in Fig.\,\ref{fig:MembraneBow}\,b). However, the effect is observed to be most pronounced for the smallest micro-lens whose colour fringes closely resemble the planar membrane device in Fig.\,\ref{fig:MembraneBow}\,a). As the lens diameter increases, the coloured fringes decrease in spatial frequency, indicating a reduced membrane bow. This can be explained by the fact that in membranes where the original layer thickness are maintained, the summation of bending moments arising from each layer are balanced. Once the thicknesses and/or volumes are altered (e.g., by partial etching of the layers to create a lens geometry), this balance is disrupted, so that the membrane starts to bow, creating an additional bending stress restoring equilibrium. This effect is here further enhanced, since lens height and the diameter are for these particular devices positively correlated, with the height rising from 2\,$\mu$m to 3\,$\mu$m, compare Fig.\,\ref{fig:BowedLensesShape} for AFM data. As more material is removed, as in the case of small radius of curvature lens structures, the bow of the devices becomes larger. Hence, the optimised micro-lenses are designed to leave more of the epitaxial stack intact by limiting the lens etch depth to 2\,$\mu$m, matching the thickness of the GaN epi. As shown in Fig.\,\ref{fig:MembraneBow}\,c), this balanced height of GaN micro-lens and AlGaN/AlN mesa did not result in air cavity induced colour fringes after the transfer step. The flatness of these membranes can be further confirmed by the tilted SEM image in Fig.\,\ref{fig:MembraneBow}\,d) which shows an upright-standing micro-lens membrane from a side-view. 

\begin{figure}[h]
\centering\includegraphics[width=0.75\linewidth]{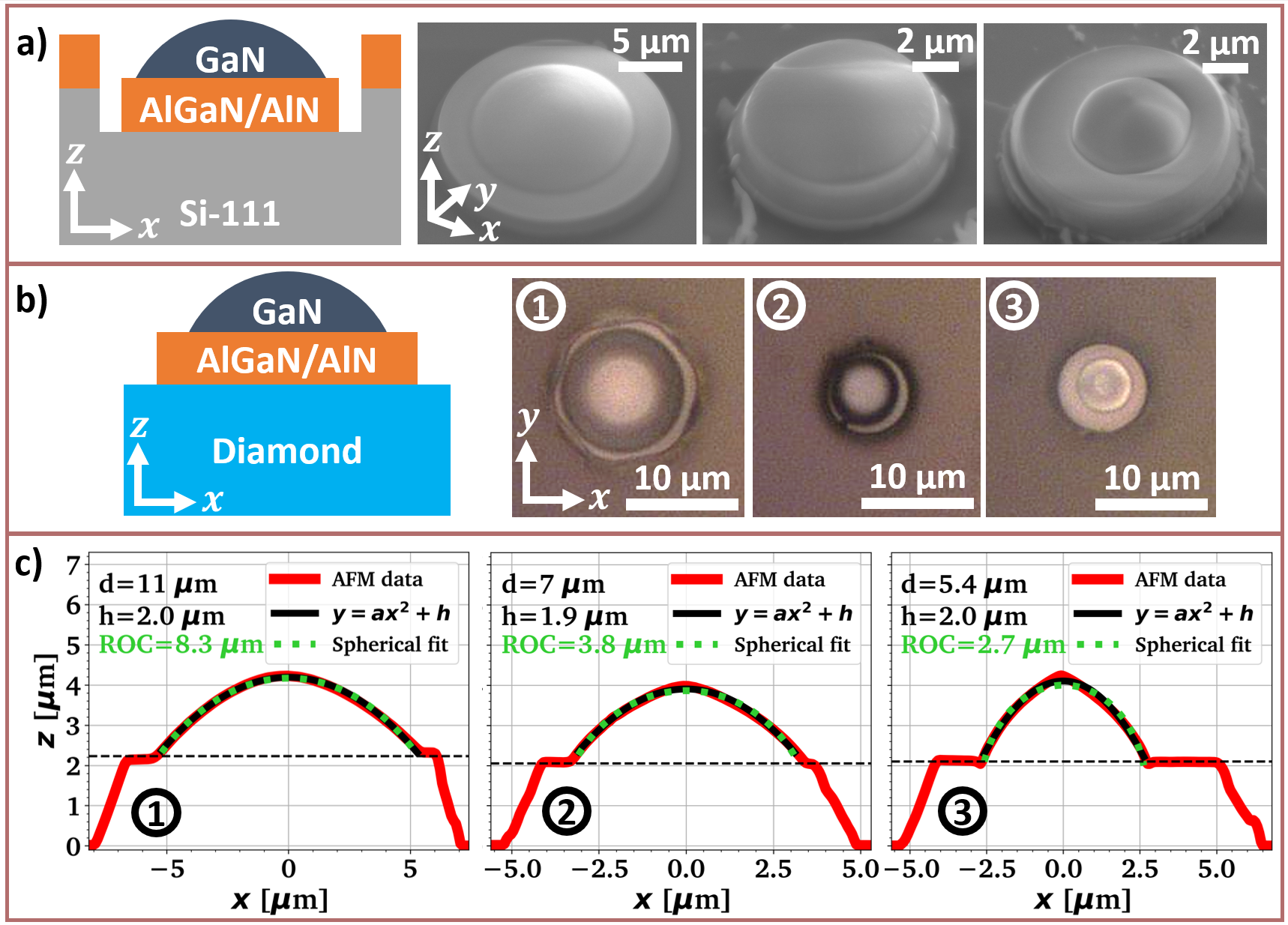}
\caption{a)\,Schematic and 40\,° tilted SEM images of collapsed micro-lenses on mesas without anchors on the donor substrate, b)\,Schematic and microscope image of the transfer-printed devices on SC diamond used for the optical analysis in Fig.\,\ref{fig:OpticalAnalysis}, c)\,AFM topography and shape analysis of the devices shown in b).}
\label{fig:ShapeAnalysis}
\end{figure}

Fig.\,\ref{fig:ShapeAnalysis}\,a) shows SEM images of devices collapsed to the Si surface after KOH etch and SiO$_x$ mask removal. For optical testing, we selected three simlar micro-lenses with varying diameter (\circled{1},\circled{2},\circled{3}) printed on a electronic grade single crystalline CVD diamond membrane from Element6 [2x2\,mm, N\,$<5$\,ppb], see Fig.\,\ref{fig:ShapeAnalysis}\,b) for microscope images. AFM profile scans of these devices are plotted in c) and fitted both with a spherical and parabolic function. Both fits match the data well, but the parabolic fit performs slightly better, showing r.m.s. deviation from the data on the order of 30-50\,nm compared to 40-110\,nm for the spherical fit. The deviations are in both cases largest for the highest aspect-ratio lens \circled{3}.

\section{Optical analysis}

\begin{figure}[h]
\centering\includegraphics[width=0.6\linewidth]{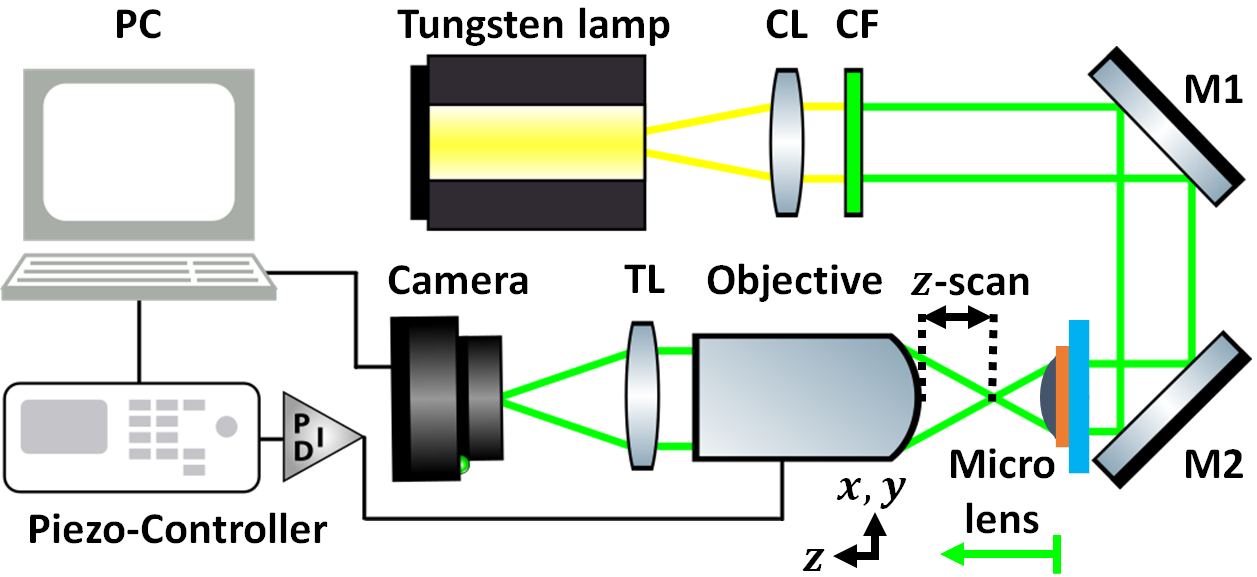}
\caption{Optical setup used to evaluate the focal length of GaN micro-lenses printed onto diamond based on\,\cite{Baranski2014}. The convex lens CL, green color filter CF, mirrors M1 and M2 and tube lens TL are annotated. Illustrations are taken from\,\cite{Franzen2022}.}
\label{fig:Setup}
\end{figure}

The focal lengths of the three lenses printed onto SC diamond substrate, shown in Fig.\,\ref{fig:ShapeAnalysis}\,b)/c), are measured using a custom built infinity-corrected microscope illustrated in Fig.\,\ref{fig:Setup}, which is based on\,\cite{Baranski2014}. The light from a tungsten lamp (Ocean optics HL-2000-FHSA) is columnated with a convex lens (CL) and passed through a color filter (CF) primarily transmitting green light. For additional spectral selection, only the green pixel channel of the CCD color camera with a Bayer filter (Allied Vision Prosilica GC650-C) is used for the analysis, leading to a total of 10\,dB rejection in the $\lambda=525\pm50$\,nm wavelength range. The measured transmission spectrum of the setup is shown in Fig.\,\ref{fig:Spectrum} in the supplementary document. The refractive index of GaN varies only by ca. $2\,\%$ in this wavelength range\,\cite{Barker1973}.\\
The collimated beam is coupled through the back-side of the diamond substrate and then through the printed GaN micro-lenses on the opposite face. The wide field illumination allows imaging of the micro-lenses with an infinity-corrected 60x objective (Nikon Plan Fluor, NA\,$=0.85$), using a 200\,mm tube lens (TL, Thorlabs LA1708-A) to image onto a CCD array. A calibrated piezo controller (PI P-725.4CD with E-665CR) is used to manipulate the objective's $z$-position with sub-micron accuracy. Automated $z$-scans are constructed by taking images in synchronisation with the piezo position.\\
The focal length $f_{air}$ of the micro-lens in air is extracted from the $z$-scan as a bright spot on the CCD camera, when the focal spot of the objective and micro-lens overlap. We then compare the $z$-travel distance between a sharp image of the micro-lens mesa and the image of the focal spot to extract $f_{air}$. The broadband source was used in these measurements to avoid obscuring interference fringes generated by monochromatic sources.

\begin{figure}%[H]
\centering\includegraphics[width=\linewidth]{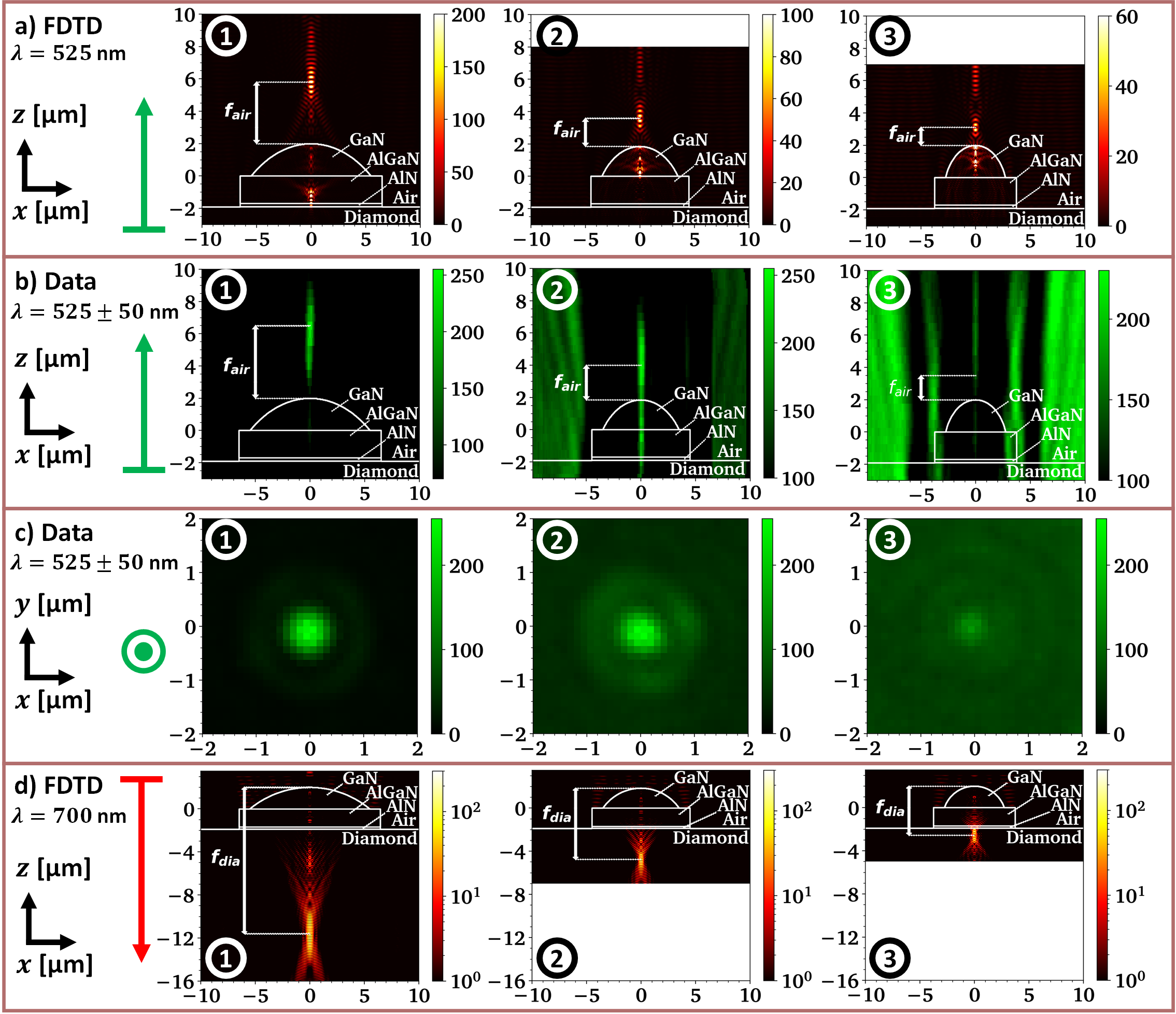}
\caption{Optical analysis of the GaN micro-lenses on diamond shown in Fig.\,\ref{fig:ShapeAnalysis}\,b)/c): a)\,FDTD simulations injecting a linearly polarised plane wave at $\lambda=525$\,nm wavelength from diamond through spherical lens profiles, b)\,Measured $z$-scan of the $x$-profile through the $xy$-CCD data evaluated on the lenses with the setup shown in Fig.\,\ref{fig:Setup}, overlayed with contours from the FDTD simulation in a), c)\,$xy$-CCD data at the objective's focal position matching $f_{air}$ in b), d)\,FDTD simulation injecting a plane wave at $\lambda=700$\,nm from air into the indicated spherical micro-lenses on diamond.}
\label{fig:OpticalAnalysis}
\end{figure}

The expected focal length of the three example micro-lenses is calculated using 3D FDTD simulations (Lumerical) in which the lenses are created to match the form factor of the measured devices, based on the AFM profiles, as shown in Fig.\,\ref{fig:ShapeAnalysis}\,c). The refractive index model for the micro-lens structure is based on \cite{Phillip1964,Barker1973, Laws2001, Pastrnak1966}, using Al$_x$Ga$_{1-x}$N with $x=38\,\%$.\\
Fig.\,\ref{fig:OpticalAnalysis}\,a) shows the simulated electric field strength squared for a plane wave injected from the diamond substrate upwards through the micro-lens geometry at $\lambda=525$\,nm wavelength. We chose the apparent symmetry point of the focal spot to evaluate the focal length $f_{air}$ measured from the GaN micro-lens tip. The results are displayed in Tab.\,\ref{tab:Results} and compared to what would be expected from the geometric optics (GO) approximation for a spherical surface with index $n_{\text{GaN},\lambda=525\,\text{nm}}=2.43$\,\cite{Barker1973}. We are evaluating the following equations\,\cite{Demtroder2009} on the parameters found in the spherical fit in Fig.\,\ref{fig:ShapeAnalysis}\,c):
\begin{equation*}
    \text{ROC}=\dfrac{\frac{d}{2}^2+h^2}{2h} \hspace{2cm}
    f_{air}= \text{ROC}\cdot \dfrac{1}{n_{\text{GaN},\lambda=525\,\text{nm}}-1}
\end{equation*}
The shortened focal lengths found in the FDTD simulations in comparison to the GO approximation shows the necessity of using a full Maxwell solver to accurately predict the micro-lens performance. The diffraction effect evident in the simulations is known as focal shift and well documented for small scale micro-optics\,\cite{Nussbaum1997,Kim2010,Lee2020}. Diffraction is also reported to account for the visible "funneling" of light into tube, rather than a distinct focal spot\,\cite{Nussbaum1997}, and can be seen in  Fig\,\ref{fig:OpticalAnalysis}\,a).\\
Fig.\,\ref{fig:OpticalAnalysis}\,b) shows the measured $z$-scan sections with 0.5$\,\mu$m step size for each micro-lens.  The lens geometries from the FDTD simulations are overlaid as a guide to the eye, with their locations defined by imaging the mesa structures. The measured position of the focal point is extracted as the highest intensity spot above the lens surface in air and is presented alongside the simulation values in Tab.\,\ref{tab:Results}. The measurement errors are dominated by the uncertainty regarding the lens surface position (found from the $z$-position of the mesa image) and the $\mu$m-large measured depth of focus in the measurement setup.\\
The $xy$-view of the selected focal spot is shown in Fig.\,\ref{fig:OpticalAnalysis}\,c). The full-width-half-maximum\,(FWHM) of the measured focal spot size is on the order of $700-800$\,nm, showing reasonable circular symmetry. We do not expect to reach the diffraction limited spot size (FWHM\,$\approx400\,$nm for $\lambda=525\,$nm and simulated NA in air of $\approx 0.67$ for all three lenses,\,\cite{Baranski2014}) as we are using a broadband beam with low coherence.

Expected and measured focal lengths agree well for lenses \circled{1} and \circled{2}, but the results for the highest-aspect ratio micro-lens \circled{3} are more challenging to extract due to the lack of a clear signature of the focal spot. This may be partially due to the reduction of contrast due to the decreasing lens aperture, as visible in Fig.\,\ref{fig:OpticalAnalysis}\,c). Additionally, spherical aberration and astigmatism might lead to a wider spread of the focus along the $z$-axis.\\
There is reasonable agreement between measured results and FDTD simulations, and so the focal length $f_{dia}$ in the diamond substrate can be estimated by inverting the FDTD simulations, see Fig.\,\ref{fig:OpticalAnalysis}\,d). In this case, a linearly polarised plane wave is injected from the top of the sample through the lens and into the substrate. Here we use a wavelength $\lambda=700$\,nm corresponding to the central wavelength of NV$^-$ emission at room temperature\,\cite{Gruber1997}. The focal length $f_{dia}$ in diamond in reference to the GaN lens tip is assessed by identifying the symmetry point of the apparent focal spot, which leads to the results presented in Tab.\,\ref{tab:Results}. These values imply that the fabricated micro-lenses could be used to couple to NV$^-$ centers both in nm-proximity to the diamond surface and in up to 10$\,\mu$m depth. To illustrate the light collection potential of our demonstrated micro-lenses, we calculate the numerical aperture for collection from diamond using the following expression:
\begin{equation*}
    \text{NA}_{dia} = n_{dia}\cdot\sin\left( \arctan\left( \dfrac{d}{2(f_{dia}-h)}\right)\right)
    %\left( \arctan(\dfrac{d}{2\cdot(f_{dia}-h}))
\end{equation*}
which indicates potentially comparable performance to high-end oil immersion objectives.

\begin{table}[h]
\centering
\caption{Results of the AFM and optical analysis of the lenses shown in Fig.\,\ref{fig:ShapeAnalysis}\,b)/c). Diameter\,$d$, height\,$h$, radius of curvature\,(ROC) and focal length\,$f$ are given for the geometric optics (GO) approximation and FDTD simulations and compared to the experimental data (EXP). $f_{air}$ is simulated at $\lambda=525$\,nm, while $f_{dia}$ is simulated at $\lambda=700$\,nm wavelength.}
\begin{tabular}{ |c|c|c|c|c|c|c|c|c| }

 \hline
   & $d$ & $h$  & ROC  & $f_{air}$ GO & $f_{air}$ FDTD  & $f_{air}$ EXP  & $f_{dia}$ FDTD & NA$_{dia}$ \\ 
  \hline
  \# & [$\mu$m] & [$\mu$m] & [$\mu$m] & [$\mu$m] & [$\mu$m] & [$\mu$m] & [$\mu$m] &  \\
 \hline
 \hline
 \circled{1} & 11 & 2.0 & 8 & 6 & 4 & $4.5\pm1.0$ & 14 & 1.0 \\ 
 \hline
 \circled{2} & 7 & 1.9 & 4 & 3 & 2 & $2.0\pm1.0$ & 7  & 1.4\\
  \hline
 \circled{3} & 5.5 & 2.0 & 3 & 2 & 1 & $1.5\pm1.0$ & 5 & 1.7 \\
  \hline
\end{tabular}
\label{tab:Index}
\label{tab:Results}
\end{table}

\section{Printed GaN micro-lens on a waveguide facet}
To show the compatibility of the free-standing micro-lens transfer with challenging device geometries, we integrated a high-aspect ratio micro-lens onto the polished facet of a straight waveguide on a GaN-on-sapphire PIC with a 600\,nm thin SiO$_2$ upper-cladding layer. The GaN/AlN waveguides have a total thickness of 1.2\,$\mu$m, with 850\,nm GaN on top of a 350\,nm AlN nucleation layer. The waveguide width is 2\,$\mu$m at the facet, but is tapered down to 1\,$\mu$m on the chip. See Fig.\,\ref{fig:Motivation}\,d) for the corresponding FDTD simulation.

\begin{figure}[h]
\centering\includegraphics[width=\linewidth]{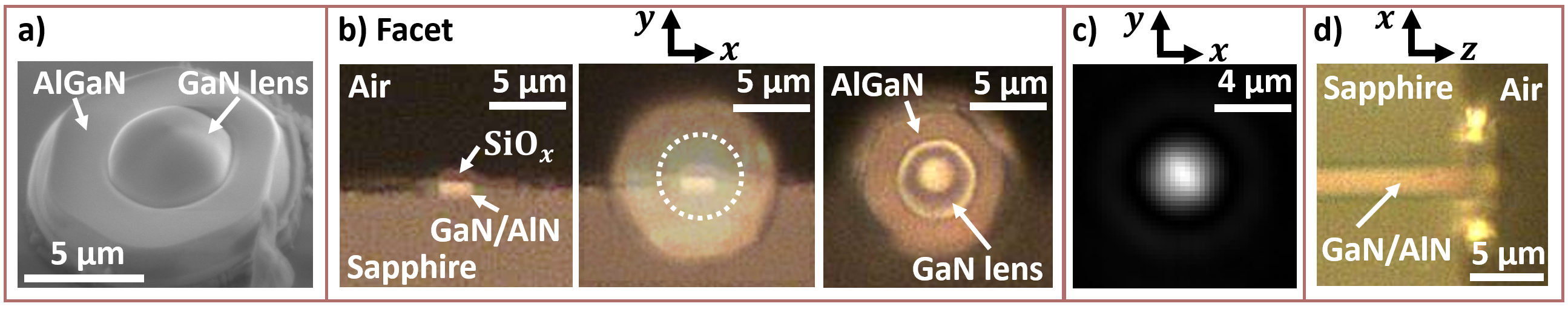}
\caption{High-aspect ratio GaN micro-lens printed on GaN-on-sapphire waveguide facet. a)\,40\,° tilted SEM image of the micro-lens on the donor sample, b)\,Microscope images of the waveguide facet (1.2x2$\,\mu$m$^2$) before (left) and after (right) transfer printing, the center image shows an overlay, c)\,Imaged output mode from the waveguide with printed micro-lens at $\lambda=1630$\,nm wavelength at 150x magnification, d)\,Top view of the printed lens after mode imaging.}
\label{fig:GaNwaveguideFacet}
\end{figure}

A tilted SEM image of the micro-lens is shown in Fig.\,\ref{fig:GaNwaveguideFacet}\,a), while microscope images of the facet before and after printing are shown in b), including an overlay of images before and after printing in the center. The chip did not contain any local alignment markers to guide a print on the facet, which restricted the alignment accuracy to around $\pm1$\,$\mu$m. The cladding layer is very thin, which causes roughly a third of the micro-lens to remain suspended in air, demonstrating the versatility of the approach and the flatness of the printed device. As previous work has shown nm-scale accuracy with transfer printing\,\cite{McPhillimy2020}, alignment to a chip with dedicated markers and a thicker cladding layer should allow printing accuracy with sub-micron precision.\\
We imaged the light output from the waveguide facet through the micro-lens using a 60x objective with a 500\,mm tube lens to reach 150x magnification. The output mode at $\lambda=1630$\,nm wavelength is shown in Fig.\,\ref{fig:GaNwaveguideFacet}\,c), indicating the transparency of the lens in this wavelength region. The top view image in Fig.\,\ref{fig:GaNwaveguideFacet}\,d) shows the lens after optical characterisation, validating the mechanical stability of the bond between lens and the polished facet even with the limited bond area. 

\section{Conclusion and Outlook}
We have demonstrated parallel fabrication of GaN micro-lenses and their deterministic dry-transfer to CVD-grown single crystalline diamond. Careful optimisation of the growth and fabrication steps enables direct contact printing of flat membrane devices onto semiconductor substrates. The etch depth of the micro-lenses strongly affects the released device flatness, highlighting the interplay between material stress profile and etch geometry. The lens imaging measurements show good agreement with FDTD simulations, which are compatible with schemes for efficient coupling to  diamond NV$^-$ centers at a range of 0.1 to 10\,$\mu$m depth from the material surface.\\
Additionally, we printed a GaN lens on a GaN/AlN waveguide facet, showing compatibility of additive micro-optics with $\mu$m-precision transfer, even when partially suspended in air. The print demonstrates both the flatness of our devices and the flexibility of the transfer-printing approach for end-fire coupling.\\ 
Future work aims to demonstrate light collection from nitrogen vacancy centres in diamond or color centres in GaN\,\cite{Berhane2017} or SiC\,\cite{Anderson2022} and efficient coupling to photonic integrated circuits in the VIS and NIR wavelength regime. Using highly optimised grayscale lithography, the benefits of free-form optics can be added to the developed additive GaN micro-optics platform. 

\appendix
\section{Backmatter}
\subsection{Funding}
Royal Academy of Engineering (Research Chairs and Senior Research
Fellowships); Engineering and Physical Sciences Research Council (EP/R03480X/1, EP/N017927/1, EP/P00945X/1); Innovate UK\,(50414);
Fraunhofer Lighthouse Project QMag
\subsection{Acknowledgement}
The authors thank Jack Smith for providing the GaN-on-sapphire photonic integrated circuit chip, which was partially fabricated in the James Watt Nanofabrication center at the University of Glasgow, Dimitars Jevtics for providing useful guidance for and help with optical measurements, Sean Bommer for 3D printing sample holders and useful discussions and the gdshelpers python package, which was used to create the layer designs of the micro-fabrication process.
\subsection{Disclosures}
The authors declare no conflicts of interest.
\section{Supplementary Document}
\counterwithin{figure}{section}
\setcounter{figure}{0}
\subsection{Additional information on process flow}
\begin{figure}[h]
\centering\includegraphics[width=0.75\linewidth]{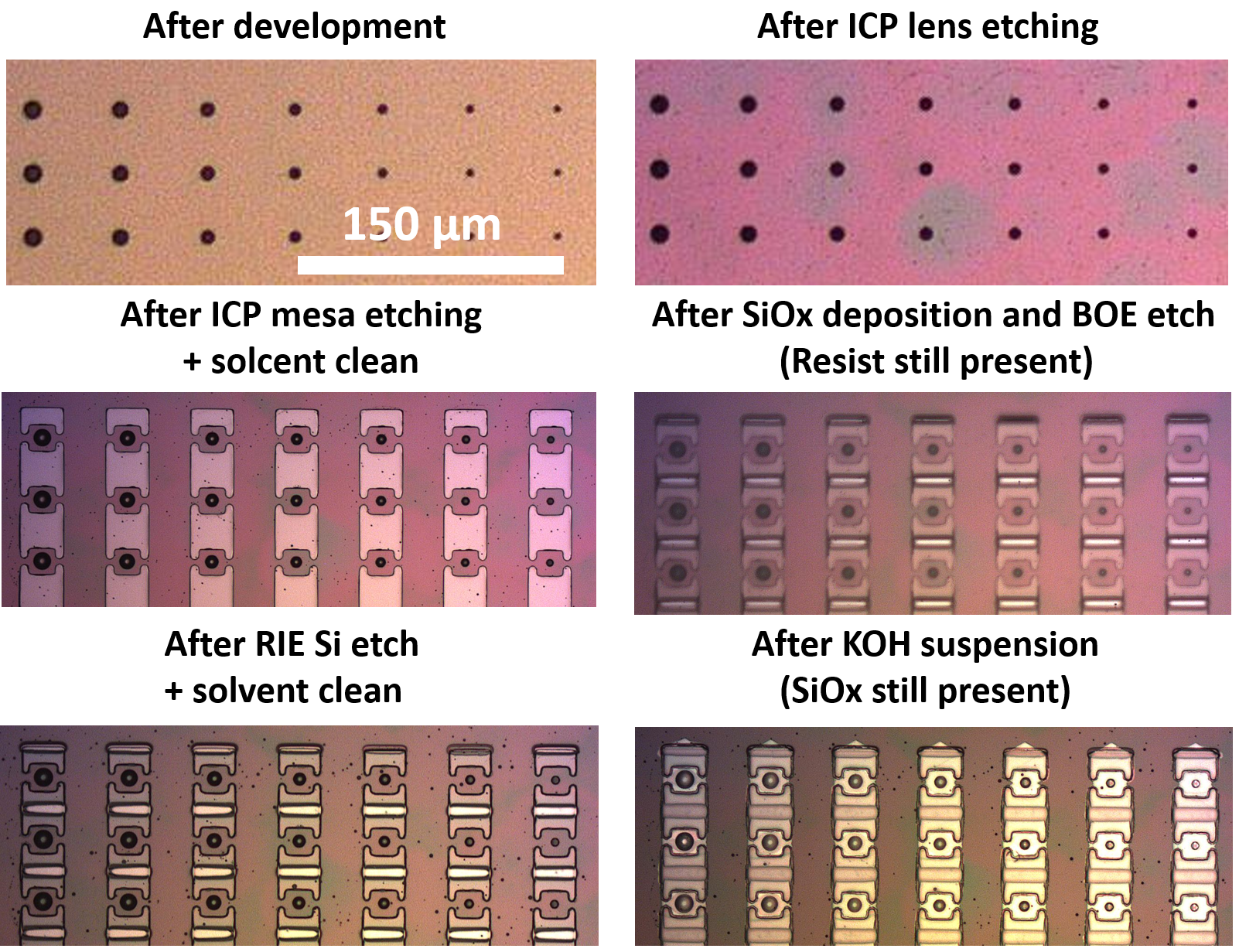}
\caption{Microscope image covering more detailed process steps for 3$\,\mu$m GaN lens etch depth.}
\label{fig:ProcessDetails}
\end{figure}

ICP-RIE GaN lens etching:\\
The employed recipe results in a comparably low 20-40 V DC bias towards the employed Si carrier wafer. The etching is done in 2.5 min steps, while purging with 50\,sccm Ar for 2 min between steps. Both low DC bias and short etching steps are found to avoid unwanted carbonisation of the top resist layer by hydrogen depletion, colloquially known as resist “burning”\,\cite{Ryu2017}.\\
Resist removal after etching:\\
Standardy solvent clean after ICP etching steps with acetone, methanol and isopropanol in ultrasonic bath.\\
Fig.\,\ref{fig:ProcessDetails} shows microscope images of the fabrication flow for 3\,$\mu$m lens etch depth for more detailed information on the etched patterns.

\subsection{AFM shape analysis of a exemplary resist and GaN lens}

\begin{figure}[H]
\centering\includegraphics[width=0.55\linewidth]{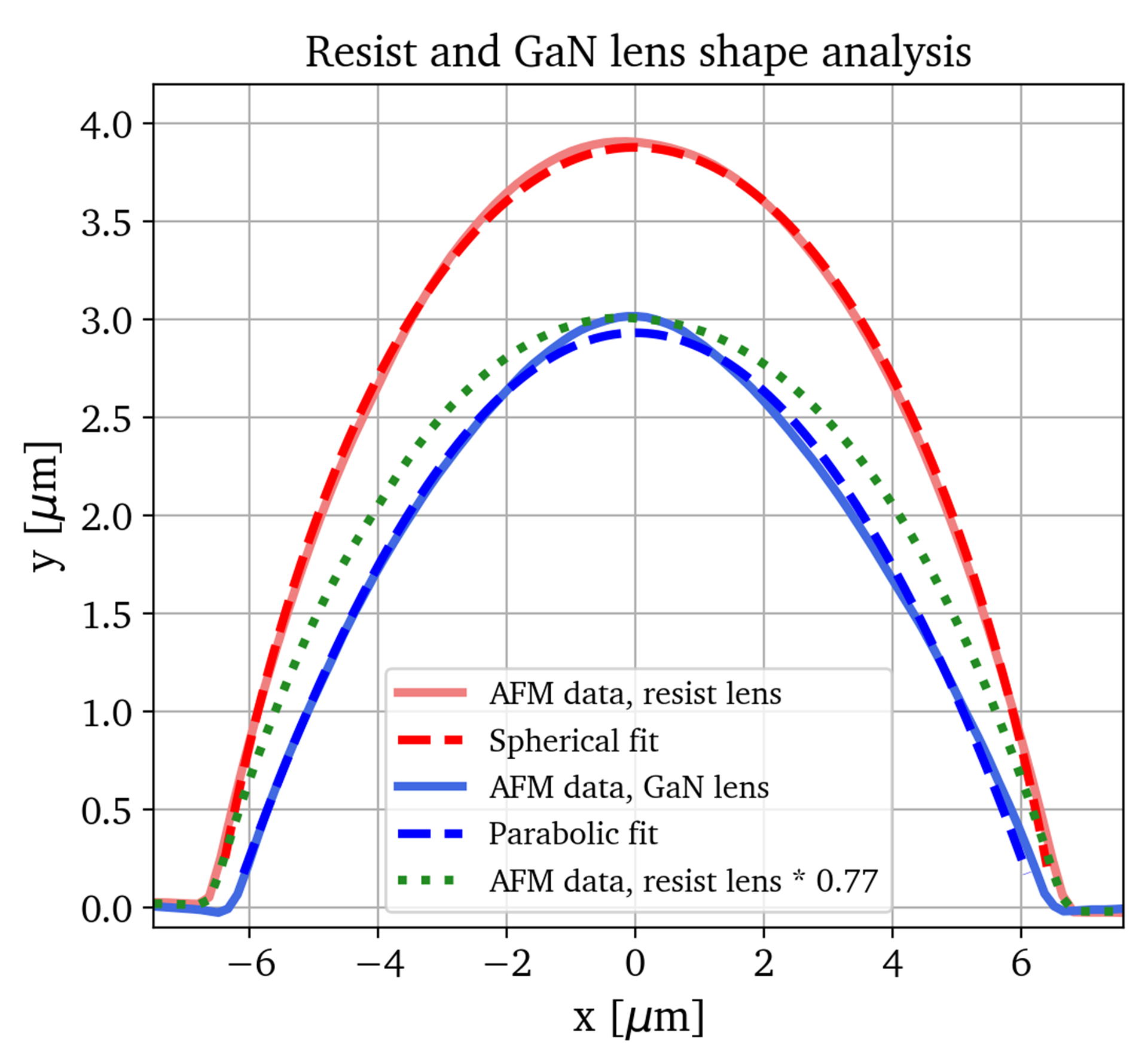}
\caption{AFM-based micro-lens shape analysis before and after ICP etching.}
\label{fig:ShapeTransformation}
\end{figure}

Fig.\,\ref{fig:ShapeTransformation} shows an example of the shape transformation that occurs between resist lens and etched GaN lens for 3\,$\mu$m etch depth and around 13\,$\mu$m diameter. The red curve shows an AFM profile of the resist lens taken after 150\,°C reflow and before etching, closely resembling the spherical fit.

The blue data shows the AFM profile of the same lens device after ICP etching and solvent clean, now closely resembling a parabolic profile. The green curve shows the expected GaN lens shape for perfectly linear etching by multiplying the resist lens profile with the selectivity $s=0.77$. The discrepancy indicates the existence of non-linear etch components.

\subsection{AFM roughness analysis of relevant surfaces}
\begin{figure}[H]
\centering\includegraphics[width=\linewidth]{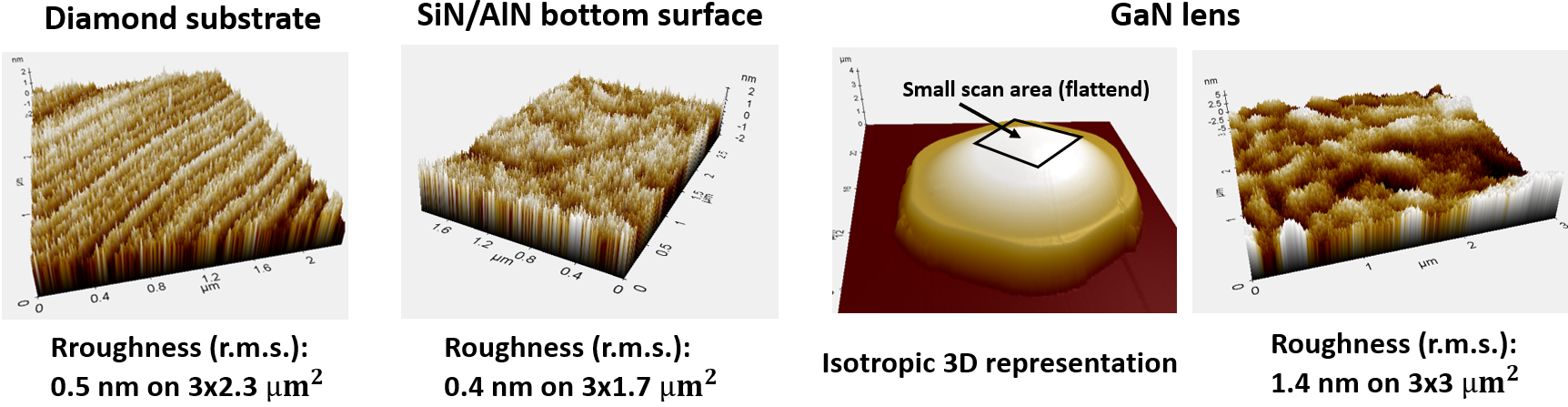}
\caption{AFM surface roughness analysis of diamond receiver substrate (polished membrane from Element 6) (left), bottom side of suspended AlGaN/AlN mesa (middle) and a GaN micro-lens printed on diamond (right).}
\label{fig:RoughnessAnalysis}
\end{figure}

Fig\,\ref{fig:RoughnessAnalysis} illustrates measured AFM surface roughness data on three relevant surfaces using 3D representations of the topography: The left hand side shows the result of a tapping mode AFM scan on the diamond receiver membrane from "Element6", the middle image contains the roughness scan on the backside of a AlGaN/AlN membrane after PDMS pick-up (measured in situ), while the right hand side shows both a large and small scale AFM scan on device \circled{1} from Fig.\,4\,b)/c) in the main text after removal of the SiO$_x$ protective layer and printing to the diamond receiver membrane.

\subsection{Yield of micro-lens suspension}
\begin{figure}%[H]
\centering\includegraphics[width=\linewidth]{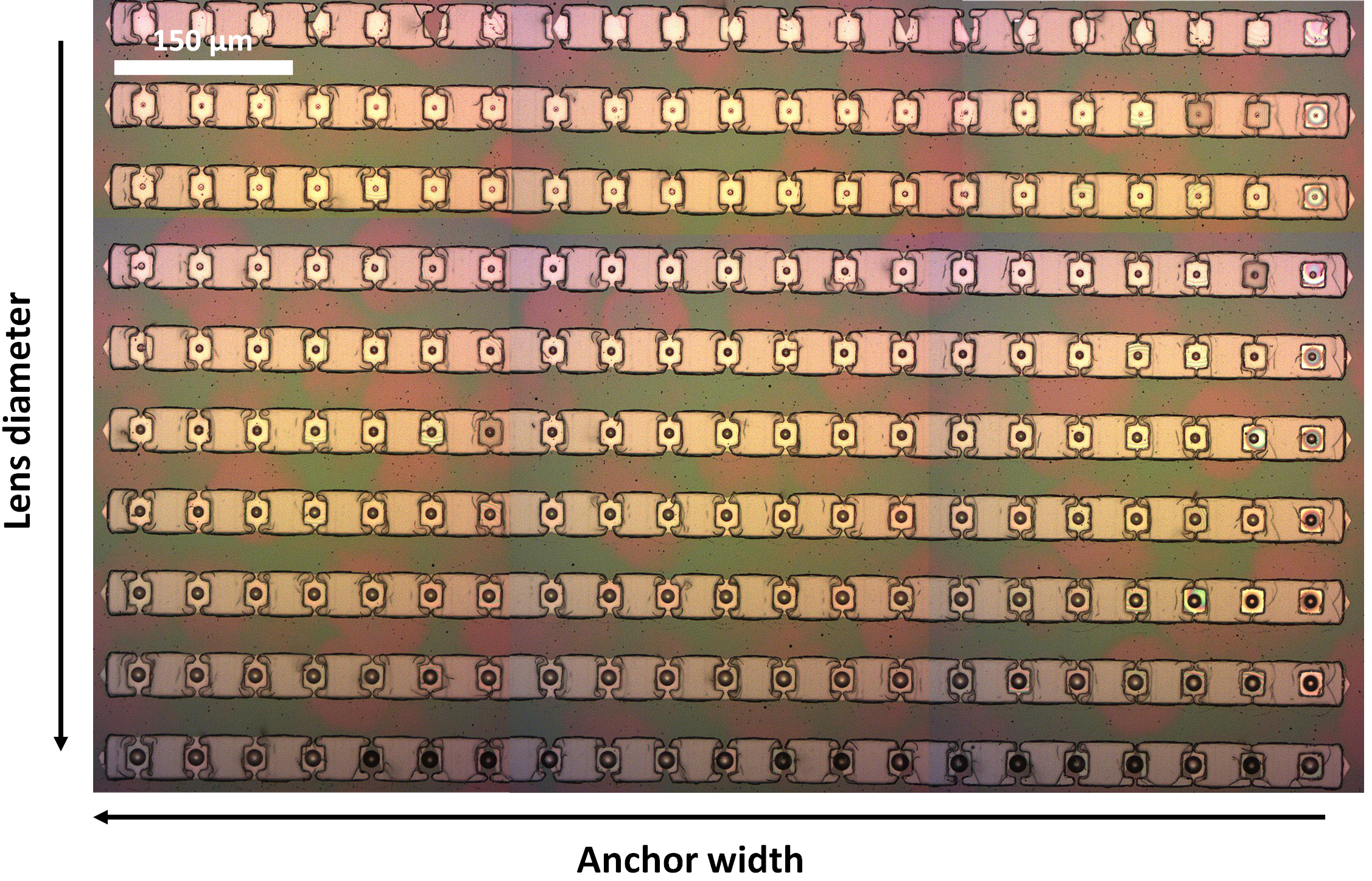}
\caption{Chip overview in the form of stitched microscope images after suspension and removal of SiO$_x$ for bowed devices on square mesas with tapered, bow-tie and straight anchor geometries.}
\label{fig:YieldofSuspension}
\end{figure}

To demonstrate the ability for large scale micro-lens production, Fig.\,\ref{fig:YieldofSuspension} shows the chip with ca. 3$\,\mu$m lens etch depth corresponding to the bowed devices shown in Fig.\,\ref{fig:MembraneBow} in the main text and Fig.\,\ref{fig:ProcessDetails}. The suspension yield can be quantified as $> 85\,\%$ for $N=210$ devices. The lens radius increases from top to bottom, while the anchor width increases from right to left.

\subsection{Dependence of membrane deflection on as-grown wafer bow}
The following Fig.\,\ref{fig:ChipletBow} illustrates the connection between wafer bow and membrane deflection for 1\,$\mu$m thick GaN epilayer on an AlGaN/AlN buffer (fully intact epilayers). The suspended membranes are still held by anchors and the minimal deflection is found on wafer C with the lowest bow.
\begin{figure}%[H]
\centering\includegraphics[width=0.9\linewidth]{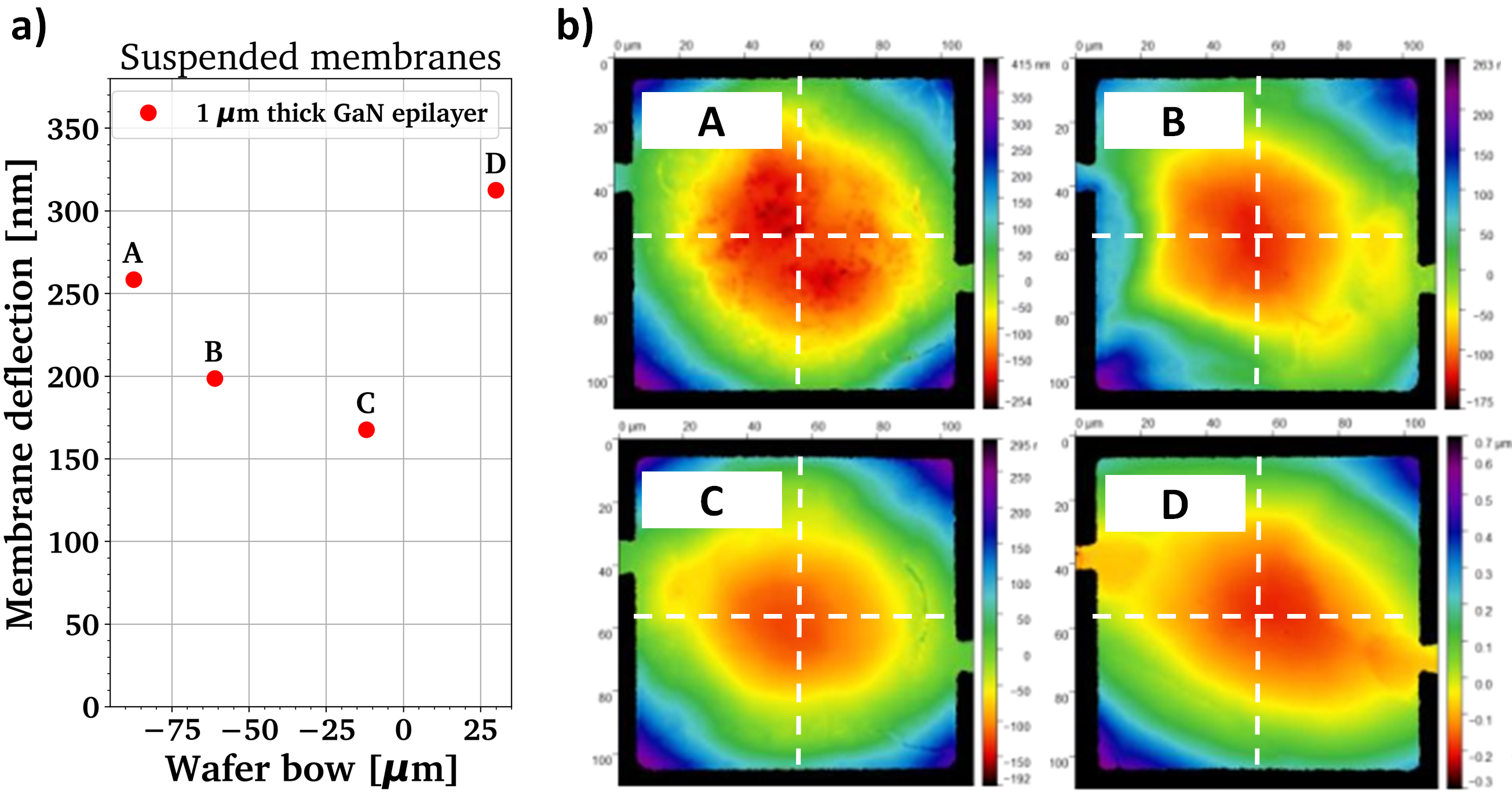}
\caption{a) As-grown wafer bow vs. deflection of suspended membranes (100x100\,$\mu$m$^2$). The averaged measured deflection from two orthogonal directions (white dashed lines in b)) of the membranes show lowest values for the wafer with smallest bow. The bow among the wafers is controlled by changing the ammonia predose time during AlN nucleation only, the mechanism of which will be published separately. b)\,Surface profilometry images of the suspended membranes from the 4 wafer runs shown in a).}
\label{fig:ChipletBow}
\end{figure}

\subsection{AFM shape analysis for bowed micro-lens devices}
The following Fig.\,\ref{fig:BowedLensesShape} shows AFM and SEM measurement results for the bowed devices with 3$\,\mu$m lens etch depth, corresponding to Fig.\,\ref{fig:ProcessDetails} and \ref{fig:YieldofSuspension}, with the AFM results explicitly belonging to the bowed devices shown in Fig.\,\ref{fig:MembraneBow} in  the main text. The lithography and reflow process yielded decreasing resist lens height with decreasing lens diameter due to simultaneous fabrication of multiple lens diameters on the same die and non-optimised grayscale preshaping. 

\begin{figure}%[H]
\centering\includegraphics[width=0.9\linewidth]{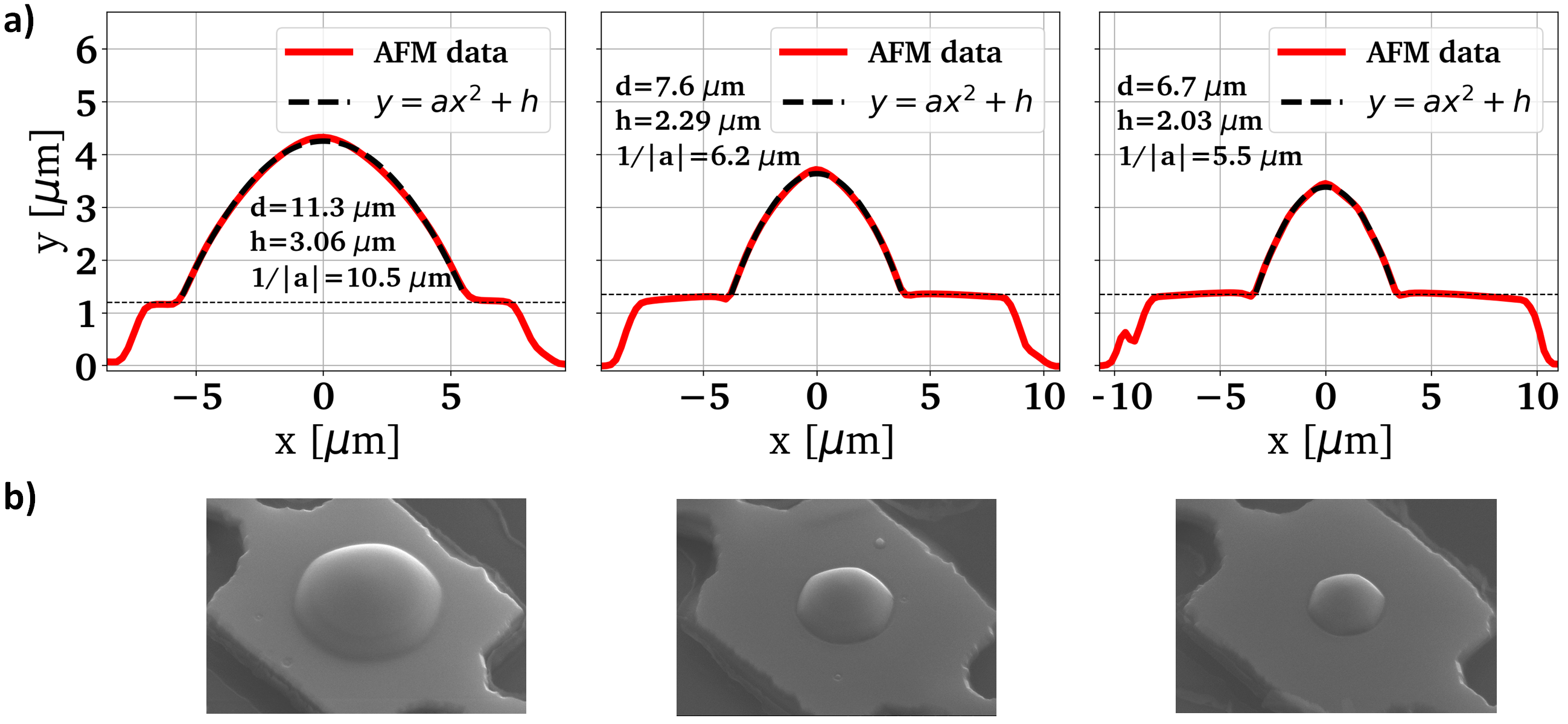}
\caption{a) AFM profiles of printed, 2-3\,$\mu$m high GaN micro-lenses on diamond with thin mesas shown in Fig.\,\ref{fig:MembraneBow}\,b) in the main text, b) SEM images corresponding to comparable devices on the same die, still on the donor substrate. Both data corresponds to the chip shown in Fig.\,\ref{fig:ProcessDetails} and\,\ref{fig:YieldofSuspension}}
\label{fig:BowedLensesShape}
\end{figure}

\subsection{Spectrum of filtered tungsten lamp used for optical analysis}
\begin{figure}%[H]
\centering\includegraphics[width=0.8\linewidth]{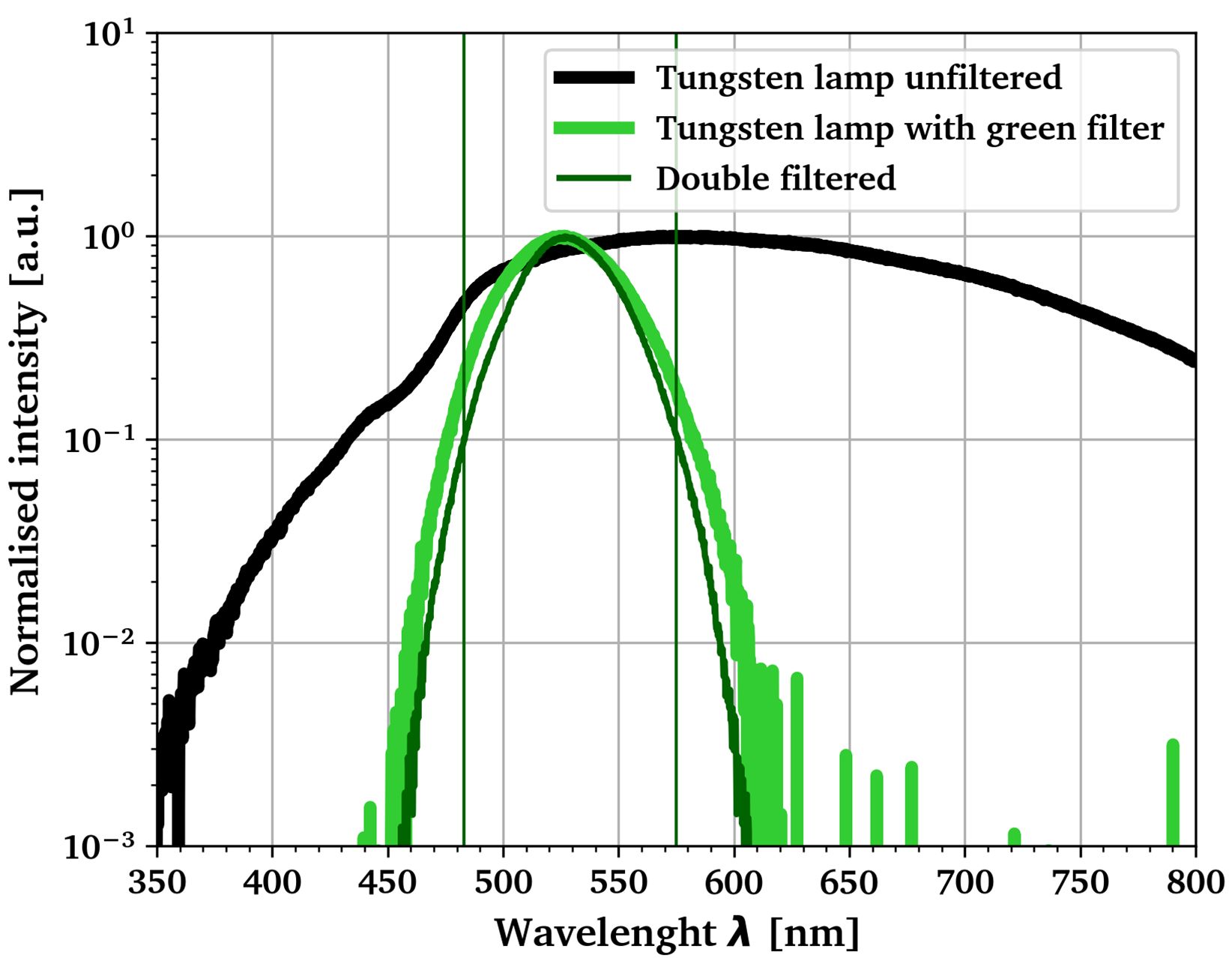}
\caption{Light source spectra used for optical analysis taken with ocean optics USB4000 fibre-coupled spectrometer (black and lime green curve) and multiplied with quantum efficiency of the green pixel of our colour camera extracted from data sheet (dark green).}
\label{fig:Spectrum}
\end{figure}

We quantified the bandwidth of the employed tungsten lamp by taking a spectrum with a fibre-coupled spectrometer, measuring the lamp unfiltered and after inserting our green color filter. We estimated the effective spectrum by multiplying the filtered spectrum with the quantum efficiency of the green pixel given by the camera data sheet, leading to a 10\,dB bandwidth in the wavelength window $\lambda=525\pm50$\,nm, as illustrated in Fig.\,\ref{fig:Spectrum}.
\newpage
\bibliographystyle{ieeetr}
\bibliography{ms}  %%% Uncomment this line and comment out the ``thebibliography'' section below to use the external .bib file (using bibtex) .

\end{document}